\documentclass[journal=jpcafh,manuscript=article]{achemso}
\usepackage{longtable}
\usepackage{multirow}
\usepackage{amsmath}
\usepackage{subfigure}
\usepackage[usenames,dvipsnames]{color}
\usepackage{xcolor}
\usepackage{soul}
\usepackage{bm}
\usepackage{rotating}

\usepackage{amssymb}
\usepackage{titlecaps}
\Addlcwords{a an the of on at with for and in into from by without}
\usepackage{soul}

\usepackage{multicol} \usepackage{wrapfig} \usepackage{enumitem}
\usepackage{bm} \usepackage{outlines} 
\SectionNumbersOn

\author{Kaisheng Song} \affiliation[University of Basel]{Department of
  Chemistry, University of Basel, Klingelbergstrasse 80, CH-4056
  Basel, Switzerland.}  \alsoaffiliation[Chongqing University]{School
  of Chemistry and Chemical Engineering, Chongqing University,
  Chongqing 401331, China.}  \altaffiliation{These authors contributed
  equally}

\author{Silvan K\"aser} \affiliation[University of Basel]{Department of
  Chemistry, University of Basel, Klingelbergstrasse 80, CH-4056
  Basel, Switzerland.}
  \altaffiliation{These authors contributed equally}

\author{Kai T\"opfer} \affiliation[University of Basel]{Department of
  Chemistry, University of Basel, Klingelbergstrasse 80, CH-4056
  Basel, Switzerland.}
  \altaffiliation{These authors contributed equally}

\author{Luis Itza Vazquez-Salazar} \affiliation[University of Basel]{Department of
  Chemistry, University of Basel, Klingelbergstrasse 80, CH-4056
  Basel, Switzerland.}

\author{Markus Meuwly} \affiliation[University of Basel]{Department of
  Chemistry, University of Basel, Klingelbergstrasse 80, CH-4056
  Basel, Switzerland.}\alsoaffiliation[Brown University]{Department of
  Chemistry, Brown University, Providence, RI,
  USA}\email{m.meuwly@unibas.ch}

\title{PhysNet Meets CHARMM: A Framework for Routine Machine Learning
  / Molecular Mechanics Simulations}

\begin{document}

\date{\today}

\begin{abstract}
Full dimensional potential energy surfaces (PESs) based on machine
learning (ML) techniques provide means for accurate and efficient
molecular simulations in the gas- and condensed-phase for various
experimental observables ranging from spectroscopy to reaction
dynamics. Here, the MLpot extension with PhysNet as the ML-based model
for a PES is introduced into the newly developed pyCHARMM API. To
illustrate conceiving, validating, refining and using a typical
workflow, para-chloro-phenol is considered as an example. The main
focus is on how to approach a concrete problem from a practical
perspective and applications to spectroscopic observables and the free
energy for the -OH torsion in solution are discussed in detail. For
the computed IR spectra in the fingerprint region the computations for
para-chloro-phenol in water are in good qualitative agreement with
experiment carried out in CCl$_4$. Also, relative intensities are
largely consistent with experimental findings. The barrier for
rotation of the -OH group increases from $\sim 3.5$ kcal/mol in the
gas phase to $\sim 4.1$ kcal/mol from simulations in water due to
favourable H-bonding interactions of the -OH group with surrounding
water molecules.
\end{abstract}

\section{Introduction}
Atomistic Simulations provide molecular-level insight into processes
and properties of chemical, biological and materials systems. This
includes both, simulations that follow the temporal evolution,
primarily based on molecular dynamics (MD), or those that sample
configurational space, i.e. Monte Carlo (MC) simulations. The
underlying object for both simulation approaches is the potential energy
as a function of the coordinates of all particles involved, which is
the full-dimensional potential energy surface (PES). With the inter-
and intra-molecular interactions described, molecular simulations 
can be carried out and experimentally
accessible observables can be determined from sufficiently long (for
MD) or extensive (for MC) calculations. This highlights the central
role PESs play for the field of atomistic simulations.\\

\noindent
Determining sufficiently accurate, full-dimensional and eventually
global PESs has remained a challenging undertaking. There are at least
three aspects that make this difficult. First, modern PESs are invariably
based on reference electronic structure calculations which are carried
out on reference geometries. The level of quantum chemical theory that
can be afforded decreases as the system size grows. In other words,
for the smallest systems, e.g. He--H$_2^+$,\cite{MM.heh2:2019} full
configuration interaction (FCI) calculations with large basis sets are
possible whereas for small peptides, e.g. tripeptides, routine
calculations for thousands of reference geometries need to resort to
lower-level density functional theory (DFT) calculations with
correspondingly smaller basis sets. Secondly, once the point-wise
reference information has been obtained, the PES needs to be
represented in one way or another as a continuous function which can
be evaluated for arbitrary geometries within the reference data
set. Such a representation should also provide accurate first
derivatives in a computationally efficient manner. Thirdly, the
representation should extrapolate reliably to regions outside those
covered by the reference data, which is often the long-range part of
the PES, and the resulting PES needs to be free of artifacts such as
``holes''.\\

\noindent
Over the past 15 years machine learning-based approaches have
flourished in addressing the points outlined above and the field has
made remarkable progress to the extent that software solutions are
becoming available for routine representation and exploration of
intra- and inter-molecular interactions. A non-exhaustive list are
methods based on permutationally invariant polynomials
(PIPs),\cite{houston2023pespip,bowman.irpc:2009} neural networks (NNs)
such as SchNet\cite{schnet:2018}, PhysNet,\cite{MM.physnet:2019} or
DeepPot-SE,\cite{zhang:2018} kernel-based methods including
(symmetrized) gradient domain machine learning
((s)GDML)\cite{chmiela:2017,sauceda:2019}, reproducing kernel Hilbert
spaces,\cite{rabitz:1996,hollebeek.annrevphychem.1999.rkhs,MM.rkhs:2017}
FCHL,\cite{fchl:2020} or Gaussian process
regression.\cite{bartok:2010,krems:2016} Several recent reviews aptly
summarize the present state-of-the
art.\cite{unke:2021,manzhos:2020,MM.cr:2021,deringer:2021}\\

\noindent
Many of the above-mentioned methods have been applied to a range of
systems (gas-phase, solids) and observables (spectroscopy, reaction
dynamics) and their usefulness has been demonstrated. Thus, the field
is ready for user-friendly implementations of the methods together
with their integration into established computer codes. The present
contribution describes the use of PhysNet together with the general
molecular simulation software CHARMM \textit{via} the pyCHARMM API and
the newly introduced MLpot module\cite{pycharmm:2023} with a focus on
implementation and workflow to illustrate how to conceive, in
principle, a robust NN-based PES for atomistic simulations. For this,
para-Cl-phenol (para-Cl-PhOH) is used as an example to highlight both,
the benefits and open issues in NN-based atomistic simulations. The
work is structured as follows: First, general aspects including data
generation, the PhysNet architecture and its training procedure, and
details concerning the ML/MM approach are outlined. Then, technical
aspects regarding the PES generation, validation and refinement for
para-Cl-PhOH are given before the results from the ML/MM simulations
are given. Lastly, the work is summarized and discussed in a broader
context.\\

\section{Technical Aspects}

\subsection{Generating a Potential Energy Surface Using PhysNet}
\begin{figure}
    \centering
    \includegraphics[width=1.0\textwidth]{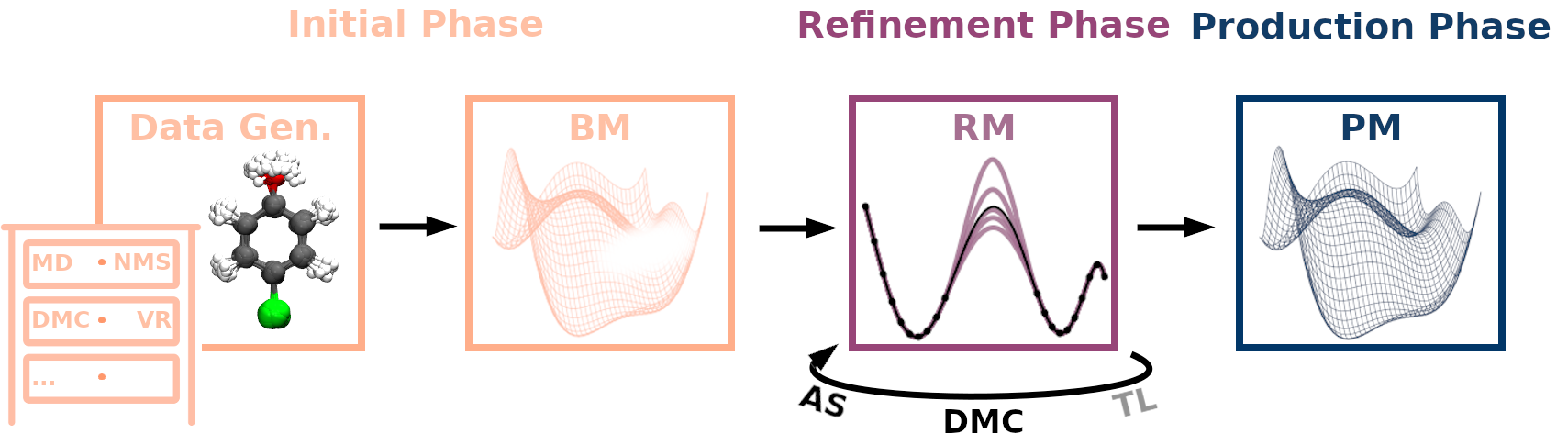}
    \caption{Schematic for the PES generation process consisting of
      three phases. The initial phase comprises the generation of an
      initial data set followed by fitting a base model. During the
      iterative refinement phase the data set is suitably extended to
      cover all (relevant) configuration space of the problem at hand
      and to detect and fill holes. This can for example be done using
      adaptive sampling and/or diffusion Monte Carlo
      (DMC).\cite{li2021diffusion} An optional step during refinement
      is to improve the PES from a lower to a higher level of quantum
      chemical rigor, for example, using TL. The production phase
      includes fitting and validating the final PES used in the production simulations for computing
      observables. Abbreviations: \textbf{MD}: Molecular dynamics
      \textbf{NMS}: Normal mode sampling \textbf{DMC}: Diffusion Monte
      Carlo \textbf{VR}: Virtual reality \textbf{AS}: Adaptive sampling
      \textbf{TL}: Transfer learning \textbf{BM}: Base model
      \textbf{RM}: Refined model \textbf{PM}: Production model.}
    \label{fig:pes_generation_scheme}
\end{figure}

Generating a NN-based PES using a machine learning approach such as
PhysNet encompasses several steps, see
Figure~\ref{fig:pes_generation_scheme}. These steps include, among
others: generation of reference data, selection of the hyperparameters
of the architecture, fitting of the model, model evaluation and
refinement.\\

\noindent
The first step is the generation of adequate reference data. There are
several strategies for this, including MD simulations 
using an empirical force field, \textit{ab initio}
MD,\cite{behler:2021} Normal Mode
Sampling\cite{ani:2017}, Diffusion Monte Carlo
simulations\cite{kosztin1996introduction,li2021diffusion}, Virtual
reality sampling\cite{amabilino2019training,amabilino2020training}, or
Atoms-in-Molecule Fragments (AMONS)
sampling\cite{lilienfeld2020amons}. The initial sampling method should
be selected with the final application in mind because each method
presents advantages and disadvantages. For example, if chemical
reactions will be studied with the final NN-based PES, generating
reference information needs to be based on a representation of the
energy function that allows bond-breaking and bond-formation, such as
{\it ab initio} MD or reactive force
fields.\cite{nagy.jctc.2014.msarmd} Conventional (biomolecular) force
fields would be unsuitable for this. For a broader discussion, see
References \citenum{unke:2021} and \citenum{MM.rev:2023}. It should
also be noted that the construction of the reference database is an
iterative process, in which the model is fitted and then refined until
it achieves the desired quality for the application at hand. \\

\noindent
The next point to consider in conceiving PESs by NNs is the
architecture of the model, i.e. the structure of the NN,
the choice of activation functions and other technical aspects. A
comprehensive description of different NN architectures can be found
in Reference \citenum{MM.rev:2023}. PhysNet, which is used in the
present contribution, belongs to the family of message-passing NNs\cite{gilmer2017neural}
which are a particular type of graph neural networks.\cite{scarselli2009gnn} 
The input to PhysNet (Figure \ref{fig:flowchart}A) 
consists of the nuclear charges and the
coordinates of all atoms of a molecule, which are then propagated
through the different layers of the model and finally converted to the
desired quantity, such as "potential energy" or "mechanical forces on
atoms". The atoms' "chemical types" are encoded into an embedding
vector as random values following a uniform distribution between
$-\sqrt{3}$ to $\sqrt{3}$ initialized by the value of the nuclear
charge. On the other hand, the coordinates of the molecule are encoded
using radial basis functions (RBFs). Those
radial functions aim at describing the {\it chemical environment} of
each of the atoms in the molecule up to a predefined cutoff. 
The embedding vector and RBFs are
then passed through $N_{\rm module}$ module blocks. Each of those
module blocks contains an interaction layer and $N_{\rm residual}^{\rm
atomic}$ residual layers. In the interaction layer, the embedding vector is
modified by combining with the RBFs and forming a message vector. The message vector is
refined by interacting with the local environment.
The  resulting output of the interaction layer is passed through
$N_{\rm res}$ residual blocks. The result is then passed through the next of $N_{\rm module}$ module blocks and the
final output layer, which converts the modified message vector to a contribution to
the atomic embedding energies $E_i$ and charges $q_i$. The contributions obtained from
the output layer for each of the $N_{\rm module}$ are summed and then
multiplied and shifted by learnable parameters depending on the atomic
charge.  Finally, the energy of the molecule of interest is obtained by
summation of the embedding energies $E_i$ of every atom predicted by
the model.
\begin{equation}
    E = \sum_{i=1}^{N}E_{i}
    \label{eq:energy_cont}
\end{equation}

\noindent
A known shortcoming of equation \ref{eq:energy_cont} is that
long-range interactions are not adequately
described\cite{MM.physnet:2019,unke:2021}. Therefore, equation
\ref{eq:energy_cont} was extended\cite{MM.physnet:2019} by including
the decay at long range using a function that smoothly damps the
Coulombic interactions for small intermolecular distances to avoid
singularities. Additional dispersion corrections to the energy, like
DFT-D3\cite{grimme2011dftd3}, can also be added. The final expression
for the energy in PhysNet is:
\begin{equation}
  E = \sum_{i=1}^{N} \left[ E_{i} + \frac{1}{2}
    \sum_{j>i}^{N}q_{i}q_{j}\cdot\chi(r_{ij}) \right] + E_\mathrm{D3}~.
  \label{eq:e_decom_physnet}
\end{equation}
$E_\mathrm{D3}$ is the DFT-D3 dispersion correction and
$\chi(\mathbf{r}_{ij})$ is a damping function defined as:
\begin{equation}
    \chi(r_{ij}) = \phi(2 r_{ij}) \dfrac{1}{\sqrt{r_{ij}^{2} + 1}} +
    (1 - \phi(2 r_{ij}))\dfrac{1}{r_{ij}}~.
\end{equation}
A continuous behaviour is ensured by the cutoff function
$\phi(r_{ij})$. \\

\noindent
Once the reference data is generated and the architecture for the NN
model is defined; the model is fitted to the data by adjusting the
weights and biases of the NN. The fitting is done by
minimizing the difference between the reference values and the values
predicted by NN model. The function which is minimised is the ``loss
function'' and can contain different quantities. For PhysNet the loss
function is
\begin{equation}
\begin{split}
    \mathcal{L} = w_{E}|E-E^{ref}| + \dfrac{w_{F}}{3N}\sum_{i=1}^{N}\sum_{\alpha}^{3}\left|-\dfrac{\partial E}{\partial r_{i,\alpha}}-F^{ref}_{i,\alpha}\right|\\
    +w_{Q}\left|\sum_{i=1}^{N}q_{i}-Q^{ref}\right| 
    + \dfrac{w_{p}}{3}\sum_{\alpha=1}^{3}\left|\sum_{i=1}^{N}q_{i}r_{i,\alpha}-p_{\alpha}^{ref}\right| + \mathcal{L}_{nh}~.
\end{split}
    \label{eq:full_lf}
\end{equation}
In equation \ref{eq:full_lf}, $E^{ref}$ and $Q^{ref}$ correspond to
the reference energy and the total charge, $F^{ref}_{i,\alpha}$ are
the Cartesian components of the force by atom, and $p_{\alpha}^{ref}$
are the Cartesian components of the reference dipole moment. The
values of $w_{i} \in \{E, F, Q, p\}$ are weighting parameters that control the
contribution of the different quantities to the loss
function. Finally, the term $\mathcal{L}_{nh}$ is a ``nonhierarchical
penalty'' that serves as a regularization to the loss
function\cite{lubbers2018hierarchical}. The loss function described in
equation \ref{eq:full_lf} is minimized by using stochastic gradient
descent techniques or variations of it. Many NN-based approaches to
representing molecular PESs use similar loss functions, however, other
possibilities can be used as well. \\

\noindent
The last step for constructing a NN-PES is the validation and
refinement of the base model. One important method to detect ``holes'' in the
PES is DMC sampling employing the base
model.\cite{li2021diffusion} Other
methods\cite{vazquez2022uncertainty} rely on the quantification of the
uncertainty of a predicted quantity (i.e. Energy and/or
Forces). Additional recommendations for validating the base model can
be found in Reference~\citenum{deringer2023validate}. Independent of the
method used to validate the generated potential, new samples will be
added to regions that were found to be under-sampled, for example by
means of adaptive sampling.\cite{csanyi2004learn} This process is
repeated until the model achieves the desired accuracy. Subsequently,
other techniques such as transfer
learning\cite{pan2009survey,smith2019approaching} (TL) or
$\Delta$-learning\cite{DeltaPaper2015} can be used to enhance the
quality of the potential, if only a limited number of ``high'' quality
reference points can be obtained.\\

\subsection{The ML/MM Approach in pyCHARMM}
One particularly promising way to use machine learned PESs is in mixed
machine learning/molecular mechanics (ML/MM) simulations. Such an
approach follows more established mixed quantum mechanical/molecular
mechanics (QM/MM) strategies in which a usually smaller part of the
system is treated with a quantum chemical method whereas the larger
remainder is represented as an empirical energy
function.\cite{cui:2021,mueller:2021,MM.fad.sol:2022,oostenbrink:2022}
In ML/MM a machine learning representation, here PhysNet, is combined
with an empirical force field such as CHARMM General Force Field
(CGenFF) as available in the CHARMM
program.\cite{Brooks.charmm:2009,cgenff2012,MM.fad.sol:2022} This
implementation into CHARMM \textit{via} the pyCHARMM API is described
next.\\

\noindent
The pyCHARMM module is a Python library that provides functions to
control the MD simulation program CHARMM via Python
commands.\cite{pycharmm:2023} It is used to combine the PhysNet PES
with the functionalities of the CHARMM program such as, e.g., the
classical force field algorithm, propagation methods and
thermostats. PhysNet computes the potential energy and forces for the
ML-atoms together with the electrostatic interactions between the
predicted fluctuating point charges of the ML-atoms and the static
atomic charges of the empirically treated
MM-atoms. CGenFF\cite{cgenff2012} handles the energies and forces for
the remaining MM-atoms and the van-der-Waals interactions between MM-
and ML-atoms. For that reason, a set of van-der-Waals parameters must
be assigned to the ML-atoms.\\

\noindent
This ML/MM approach is comparable with the mechanical embedding scheme
of the QM/MM approach, where the intramolecular energy and charge
distribution of the ML- or QM-atom system is not affected by the
presence of the surrounding ML atoms and their partial atomic charges.
\cite{mennucci:2020,kato:2018} The advantage of the current mechanical
embedding scheme is the direct application of ML-based models of
atomic systems trained in the gas-phase for condensed-phase
simulations without additional training.\\

\begin{figure}
    \centering
    \includegraphics[width=1.0\textwidth]{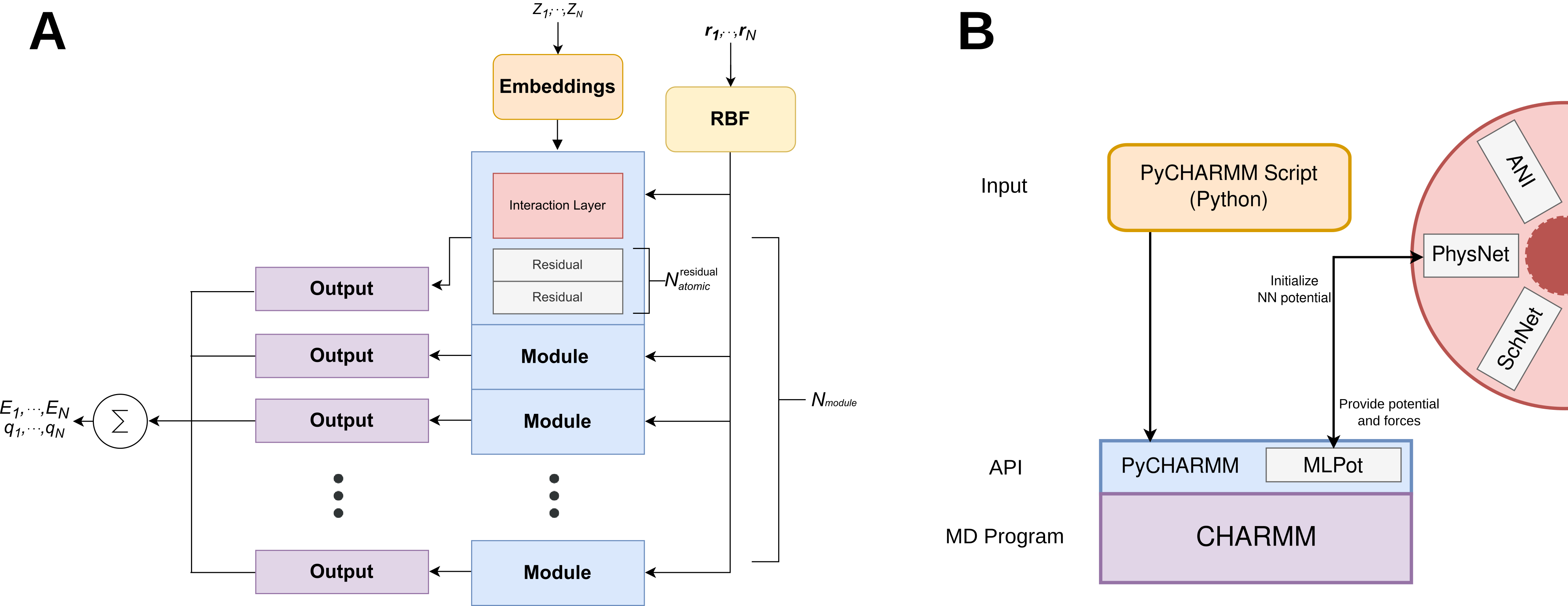}
    \caption{Panel A: Overview of the PhysNet architecture. The inputs
      to the model are the atomic numbers which create the 'embedding'
      of the different atoms and the atomic positions that are encoded
      through radial basis functions (RBF). Those two elements are then
      passed through $N_{\rm module}$ modules. Each module contains an
      interaction layer and $N_{\rm atomic}^{\rm residual}$ residual
      layers. The output of the module is passed through an output
      layer which returns an atomic energy and charge, which are
      summed to obtain the final energy and total charge of the
      molecule. Panel B: Schematic for initializing and integrating
      the PhysNet PES in simulations with the CHARMM code.}
    \label{fig:flowchart}
\end{figure}

\noindent
Figure \ref{fig:flowchart}B shows the modules which are involved in
performing MD simulations in CHARMM using PhysNet. The input file to
run pyCHARMM is a Python script and additional parameter and
topology files are required from the user to define the atomic system,
force field parameters and instructions for performing simulations
based on CHARMM. The pyCHARMM module/interface evaluates and
translates the commands to CHARMM-compatible
instructions.\cite{pycharmm:2023} The new MLPot module initializes an
external model potential and evaluates potential energy and forces for
the subset of ML-atoms together with the CHARMM force field energy. By
adapting the MLPot module in the source code, it is possible to link
different model potentials such as ANI\cite{ani:2017} or
SchNet\cite{schnet:2018}. The requirements for the model potentials
are to provide potential energy and forces. If the ML-based PES does
not predict atomic charges, the electrostatic contribution between
assigned static point charges of the ML- and MM-atoms are computed by
the empirical energy function.\\

\section{The PES for Para-Chloro-Phenol}
This section describes the generation and testing of a
full-dimensional, non-reactive PES for para-Cl-PhOH based on PhysNet.
First, the data generation is described, which is followed by the
training and testing of the base models (B$_1$ to B$_N$), their
refinement (R$_1$ to R$_M$) and transfer learning (TL$_1$ to
TL$_K$). Here, $N$, $M$, and $K$ are independently trained models
based on the same reference data set which increases in size from base
to refined models whereas the data set for TL is
comparatively small as the case for training from a lower level theory
(MP2/6-31G(d,p)) to a higher level (MP2/aug-cc-pVTZ) is considered
here.\\

\subsection{Generation of the Reference Data}
First, an initial data set needs to be generated. In the present case,
this was done by running finite-temperature simulations using the
semi-empirical tight binding GFN2-xTB method
\cite{bannwarth2019gfn2}. Simulations were run at 2000 K, 2500 K, and
3000 K, respectively and $\sim 1000$ reference structures per
temperature were extracted at regular intervals. For all these
structures, potential energies, forces and dipole moments on all atoms
were determined at the MP2/6-31G(d,p) level of theory using
MOLPRO2020\cite{MOLPRO}. Structures for which the single-point
calculations did not converge were discarded without further
analysis. All other structures were used in training 4 independent
base models using energies, forces and dipole moments.\\

\subsection{Training and Validation of the Base Model}
Training of a NN-based model requires to divide the available data
into ``training'', ``validation'', and ``test'' sets. For the present
work, a split of $70/10/20~\%$ (often also $80/10/10~\%$) was
used. After initialization, the set of optimal parameters of
PhysNet are determined by minimizing the loss function $\mathcal{L}$
(Equation~\ref{eq:full_lf}) using AMSGrad\cite{reddi2019amsgrad} and a 
learning rate of $10^{-3}$. The batch size, which corresponds to the number
of randomly chosen reference structures for which the loss is calculated at
once, was set to 32. The training was terminated upon convergence (or even
re-increasing due to overfitting) of the validation loss.\\

\begin{figure}
    \centering
    \includegraphics[width=\textwidth]{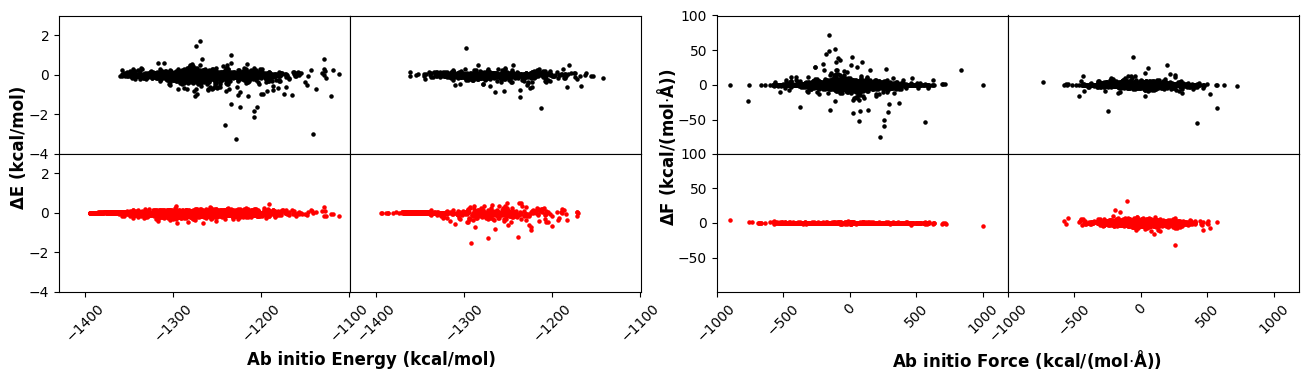}
    \caption{Comparison between reference MP2/6-31G(d,p)
      energies/forces and predicted energies/forces in the training
      and test sets, from left to right. The performance of the base
      and refined models for para-Cl-PhOH are shown in the upper and
      lower panels, respectively. Here, $\Delta E = E_{\rm PhysNet} - E_{\rm MP2}$, $\Delta
      F = F^{\alpha}_{\rm PhysNet} - F^{\alpha}_{\rm MP2}$ where
      $\alpha = (x,y,z)$ are the three Cartesian components of the
      forces on each atom. On the energies for the base model, the MAE$_{\rm train}(E)$ and MAE$_{\rm test}(E)$ are 0.11, 0.10 kcal/mol, and the corresponding RMSE$_{\rm train}(E)$ and RMSE$_{\rm test}(E)$ are 0.22, 0.18 kcal/mol. The MAE$_{\rm train}(F)$ and MAE$_{\rm test}(F)$ on forces for the base model are 0.28, 0.27 kcal/(mol$\cdot$\AA), and the corresponding RMSE$_{\rm train}(F)$ and RMSE$_{\rm test}(F)$ are 1.2,
      0.89 kcal/(mol$\cdot$\AA). Similarly, the MAE$_{\rm train}(E)$ and MAE$_{\rm test}(E)$ on the energies for the refined model are 0.04, 0.07 kcal/mol, and the corresponding RMSE$_{\rm train}(E)$ and RMSE$_{\rm test}(E)$ are 0.06, 0.16 kcal/mol. The MAE$_{\rm train}(F)$ and MAE$_{\rm test}(F)$ on forces for the refined model are 0.03, 0.32 kcal/(mol$\cdot$\AA),  and the corresponding RMSE$_{\rm train}(F)$ and RMSE$_{\rm test}(F)$ are 0.05, 0.80 kcal/(mol$\cdot$\AA).}
    \label{fig:fittingerror}
\end{figure}

\noindent
The performance of the best base model on energies and forces is
reported in the top row of Figure \ref{fig:fittingerror}. Here, the
size of the training set was 2200 energies, corresponding forces, and
dipole moments, whereas the test set contained 795 data points. Over a
range of $\sim 250$ kcal/mol ($\sim 10$ eV) the MAE$(E)$ and RMSE$(E)$
on the energies are 0.1 and 0.2 kcal/mol, respectively whereas they
are 0.28 and 1.2 kcal/(mol$\cdot$\AA) on forces. For the test
set the performance appears even better on average. It is also seen
that training and test sets cover a comparable range of energies and
forces. In line with the findings for the energies, the deviations in
the forces (third and fourth column in Figure \ref{fig:fittingerror})
are smaller for the test set than for the training set.\\

\begin{figure}[!htb]
    \centering
    \includegraphics[width=\textwidth]{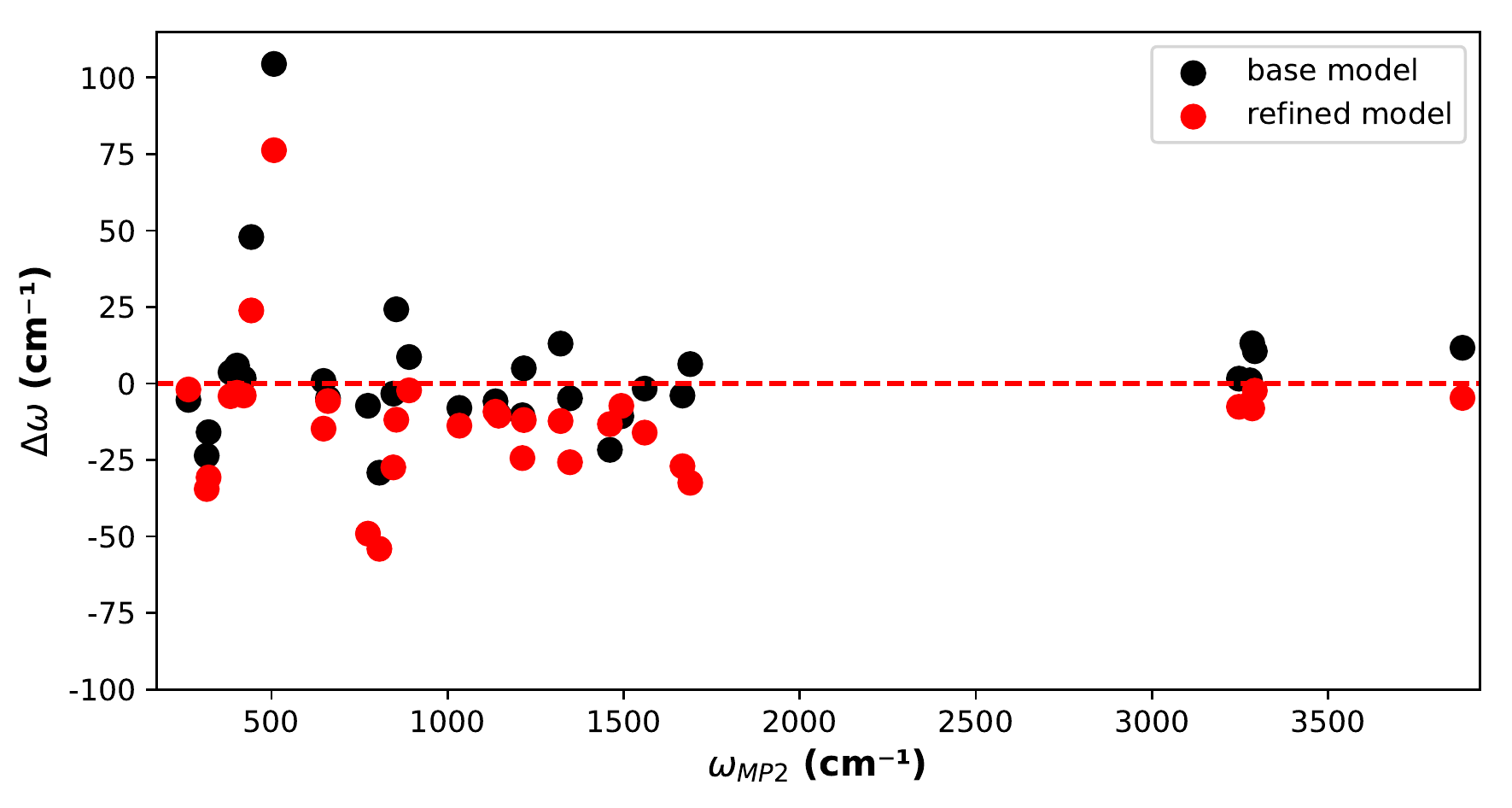}
    \caption{Accuracy of the harmonic frequencies of the base model
      (black dots) and refined model (red dots) with respect to the
      reference MP2/6-31G(d,p) values. Here, $\Delta\omega$
      corresponds to $\omega_{\rm PhysNet} - \omega_{\rm MP2}$. For the
      refined model, most of the absolute deviations are obviously
      smaller.}
    \label{fig:freq}
\end{figure}

\noindent
As a preliminary validation of the base model, minimum energy
structures and harmonic frequencies were determined from the PhysNet
representation and at the MP2/6-31G(d,p) level for comparison. The
root mean squared difference between the two optimized structures is
0.004\,\AA. This suggests that the minimum energy structure at this
level of theory is reliably captured although it was not explicitly
included in the training data set. Further improvements can therefore
be gained by including such dedicated information either in the
training set for the base model or for the refined models, see
below. The difference between reference normal mode frequencies and
those obtained from the base models is illustrated as black circles in
Figure~\ref{fig:freq}. The overall MAE and RMSE$(\omega)$ are 13 and
23 cm$^{-1}$, respectively, with a maximum deviation of slightly above
100 cm$^{-1}$. To put this into perspective, this compares to a
performance of PhysNet PESs for, e.g. ten atom formic acid
dimer\cite{MM.fad:2022} (nine atom
malonaldehyde\cite{MM.ht:2020,kaeser2023microsecond}) with
MAE($\omega$) = 2.1 ($<5$)~cm$^{-1}$ and a maximum deviation of $\sim
7$ ($< 20$)~cm$^{-1}$. These results, however, are derived from
production models that were obtained after a refinement phase with
multiple rounds of adaptive sampling. Additionally, including
structures obtained from normal mode sampling\cite{ani:2017} or
information along specific modes is likely to further improve the
performance of the models. \\

\begin{figure}[!ht]
    \centering
    \includegraphics[width=0.8\textwidth]{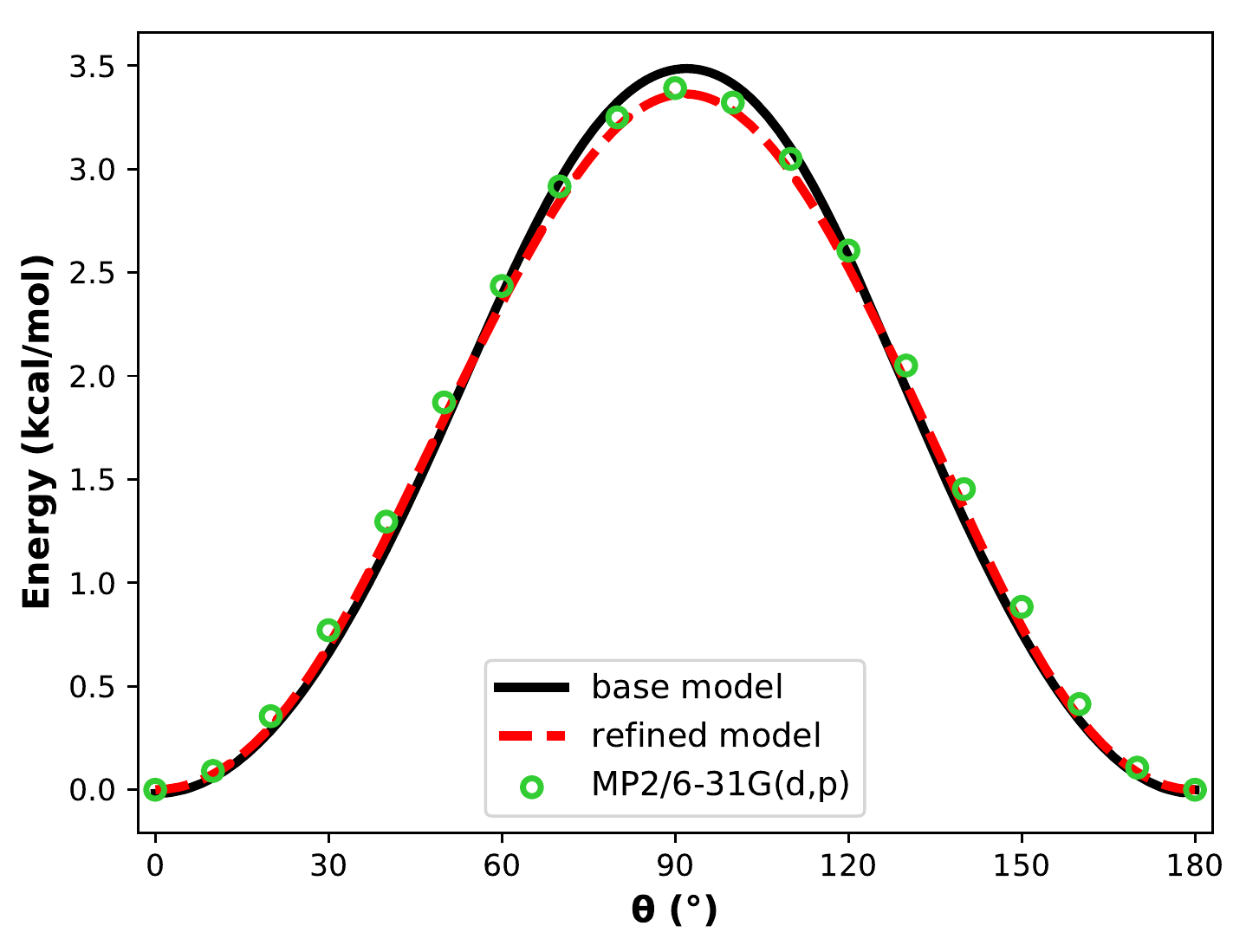}
    \caption{The energy profile of the OH torsion in
      para-Cl-PhOH. Here, the black solid line and red dashed line
      stand for the energies from the base and refined PhysNet
      models, respectively. The green open circles represent the
      reference MP2/6-31G(d,p) energies.}
    \label{fig:scan-rot}
\end{figure}

\noindent
Similarly, the base model was validated by computing the relaxed
torsional potential for the OH rotation from both, the NN-model and at
the MP2/6-31G(d,p) level of theory, see Figure \ref{fig:scan-rot}. The
green open circles are reference calculations at the MP2/6-31G(d,p)
level whereas the black solid line is from one of the base
models. Consistent with the molecular symmetry of the molecule, the
two equivalent minima are isoenergetic. Except for the region around
the transition state the base model closely follows the minimum energy
path which was, however, not yet explicitly included in the training
set. The absolute difference in the barrier height is 0.09 kcal/mol
for a barrier of $\sim 3.5$ kcal/mol. \\

\subsection{Refinement and Validation Using Adaptive Sampling}
Usually, a single base model is not sufficient for a robust PES to be
used in molecular simulations. Consequently, the model and the
underlying data set need to be extended and improved by adding
training samples. This can be done in various ways and the procedure
opted for may also depend on the application(s) one has in mind. One
established method uses adaptive sampling.\cite{csanyi2004learn} The
protocol requires two or multiple independently trained base models
which can employ different random seeds and/or reference data for
training them. With base models B$_1$ to B$_N$ finite-temperature
simulations are carried out and structures are saved if the two or
multiple base models differ in energy by a given threshold. In other
words, the disagreement in the predictions of B$_1$ to B$_N$ is used
as the criterion for adding structures together with their energies
and forces to the training set.\\

\noindent
In the present case, 1000, 1500 and 1000 additional structures with an
energy difference threshold of 0.5 kcal/mol were generated from
finite-temperature simulations at 2000, 2500 and 3000 K,
respectively. For these structures, MP2/6-31G(d,p) calculations were
carried out and the energies and forces were added to the pool of
reference data. This yielded a total of 6748 data points which were
split according to $80:10:10~\%$ into training, validation and test
samples. Next, two independent models based on the refined data set
were trained. The fitting errors on the training and test data are
reported in the bottom row of Figure \ref{fig:fittingerror}. Now, the
errors on the training set are considerably lower than those on the
test set for both, energies and forces. It is also seen that adaptive
sampling adds low-energy structures which extend the energy range
covered to close to 300 kcal/mol.\\

\noindent
Validation of the refined models proceeds along the same lines as for
the base model. The minimum energy structures now differ from the
reference MP2/6-31G(d,p) geometries by 0.0026 \AA\/. For the harmonic
frequencies, the MAE and RMSE$(\omega)$ change to 18 and 24 cm$^{-1}$,
respectively, i.e. both the MAE and RMSE slightly increase in
comparison with the base model. Specifically, the high-frequency modes
improve whereas for the low-frequency modes the average performance
remains similar with the exception of the largest outlier which
improves by 25 \%, see Figure \ref{fig:freq}.\\

\noindent
Finally, the OH-torsion profile was recomputed for one of the refined
models. The barrier height now differs by only 0.03 kcal/mol, which is
the main improvement compared with the base model. The results in
Figure \ref{fig:scan-rot} show that adaptive sampling improves the
barrier height without, however, explicitly including this information
in the training set.\\

\subsection{Transfer Learning}
To further improve the quality of the PES,
TL\cite{smith2019approaching,pan2009survey,taylor2009transfer} from
the MP2/6-31G(d,p) to the MP2/aug-cc-pVTZ level of theory was
performed. TL builds on the knowledge acquired by solving one task
(representing the MP2/6-31G(d,p) PES) to solve a new, related task
(representing the MP2/aug-cc-pVTZ PES).\cite{pan2009survey} TL and
related approaches including,
e.g. $\Delta$-learning\cite{DeltaPaper2015} or hierarchical
ML\cite{dral2020hierarchical}, gained a lot of attention in the field,
as it allows to reach system sizes which are difficult to reach with
conventional
approaches.\cite{nguyen1995dual,nandi2019using,dral2020hierarchical,kaser2021transfer,bowman2022delta,kaser2022transfer}
Here, the data set for TL contained 338 geometries chosen from the
data set for the refined model. Besides, additional 361 geometries
along the MEP of OH torsion were included to improve the performance of the model in the torsion MEP. \textit{Ab initio} energies, forces and dipole
moments for the 699 para-Cl-PhOH geometries were determined at the
MP2/aug-cc-pVTZ level of theory using MOLPRO\cite{MOLPRO}. Then, the
PhysNet model was retrained on the TL-data set by initializing the NN
with the parameters from one of the refined models at the
MP2/6-31G(d,p) level as an initial guess. The data set was split
randomly according to 80/10/10~\% into training/validation/test set
and the learning rate was reduced to $10^{-4}$ compared with $10^{-3}$
when learning a model from scratch.\\

\begin{figure}
    \centering \includegraphics[width=\textwidth]{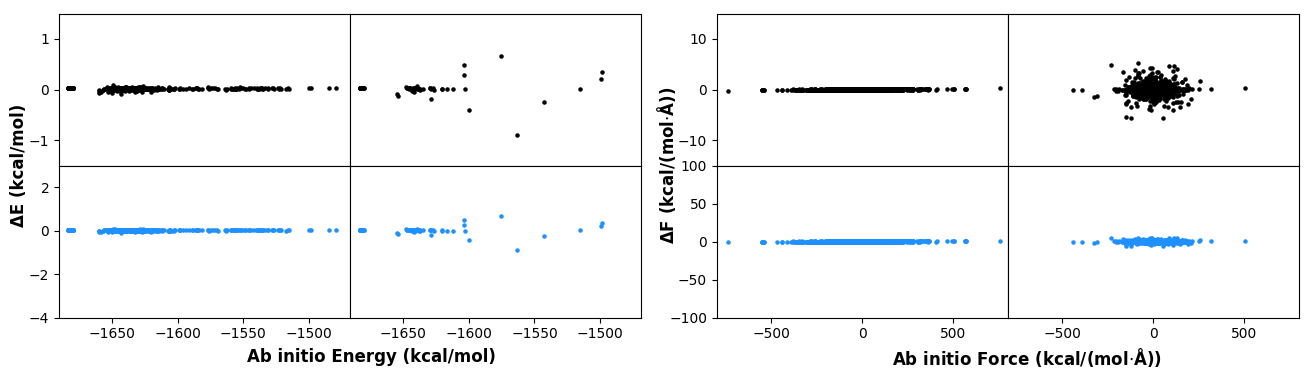}
    \caption{Comparison between reference MP2/aug-cc-pVTZ
      energies/forces and predicted energies/forces in the training
      and test sets, from left to right. The performance of the TL
      PhysNet model with the same scale along the $y-$axis as in
      Figure \ref{fig:fittingerror} is reported in the lower row and
      whereas the upper row has enlarged scales along the $y-$axis. The errors in the forces
      on the test set are considerably reduced compared with the base
      and refined models at MP2/6-31G(d,p) levels. Here, $\Delta E =
      E_{\rm PhysNet} - E_{\rm MP2}$, $\Delta F = F^{\alpha}_{\rm
        PhysNet} - F^{\alpha}_{\rm MP2}$ where $\alpha = (x,y,z)$ are
      the three Cartesian components of the forces on each
      atom. The MAE$_{\rm train}(E)$ and MAE$_{\rm test}(E)$ on the energies are 0.02, 0.8 kcal/mol, and the corresponding
      RMSE$_{\rm train}(E)$ and RMSE$_{\rm test}(E)$ are 0.02, 0.17 kcal/mol. On forces, the MAE$_{\rm train}(F)$ and MAE$_{\rm test}(F)$ are 0.01, 0.22 kcal/(mol$\cdot$\AA), and the corresponding RMSE$_{\rm train}(F)$ and RMSE$_{\rm test}(F)$ are 0.01, 0.6 kcal/(mol$\cdot$\AA).}
    \label{fig:tf_errors}
\end{figure}

\noindent
The performance on the TL training and test sets for energies and forces
is reported in Figure~\ref{fig:tf_errors}. The training data covers
$\sim 200$ kcal/mol which is a somewhat narrower range than for the
lower level model for which it was $\sim 250$ kcal/mol. For energies
(columns 1 and 2 in Figure \ref{fig:tf_errors}) the absolute error is
lower than 0.1 kcal/mol throughout on the training set and never
exceeds $\pm 1$ kcal/mol on the test set. This is a considerably
improved performance compared with the base models. Similarly, for
forces (columns 3 and 4) the errors are much smaller than for base and
refined models, see Figure \ref{fig:fittingerror}. Importantly, TL
requires only 10 \% of the training effort in time compared with
training base and refined models. This is seen in that the lowest
value of the loss function is achieved after $\sim 10^5$ iterations
whereas base and refined models were trained for $\sim 10^6$
iterations.\\

\noindent
Testing of the transfer learned models required computation of
reference data (harmonic frequencies, OH torsional barriers) at the
MP2/aug-cc-pVTZ level for direct comparison with predictions from
PhysNet. Harmonic frequencies at the MP2/aug-cc-pVTZ level of theory
for a molecule the size of para-Cl-PhOH already become prohibitive in
terms of computing time and memory requirements whereas a few hundred
reference calculations for energies and forces at the same level of
theory can be carried out in parallel and quite
efficiently. Comparison of the harmonic frequencies with their
\textit{ab initio} reference, shows that the transfer learned model
predicts the values with a MAE of $\sim 16$ cm$^{-1}$ and with maximum
absolute deviations of less than 100 cm$^{-1}$. It is expected that
including judiciously chosen structures in the TL data set drastically
reduces the deviations for the PhysNet predictions. This has been
demonstrated recently for malonaldehyde \cite{kaeser2023microsecond}
for which TL using 50 to 100 CCSD(T) quality data points were
sufficient to reach MAE($\omega$) $< 5$~cm$^{-1}$.\\

\begin{figure}
    \centering \includegraphics[width=0.8\textwidth]{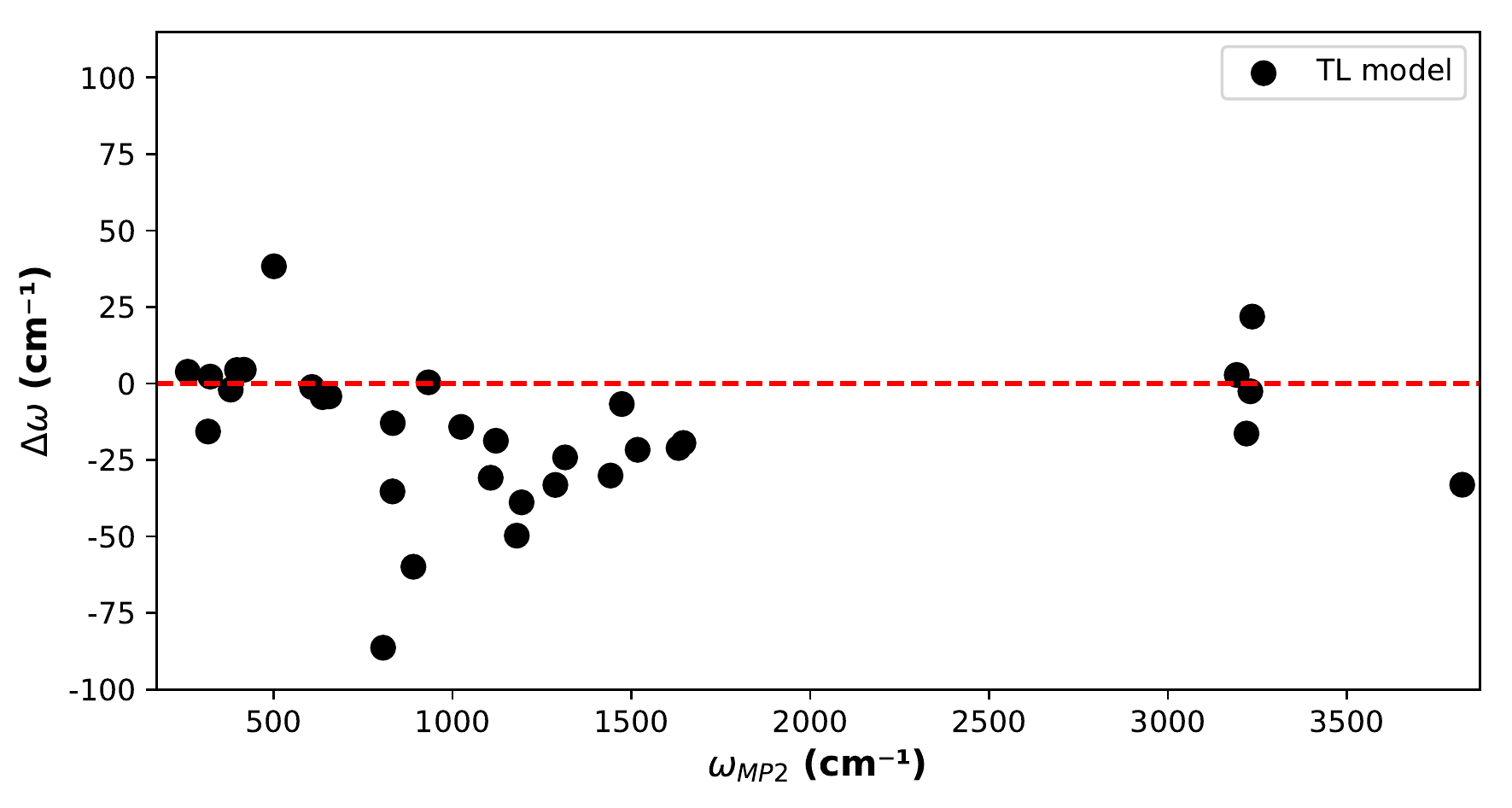}
    \caption{Transfer Learning to MP2/aug-cc-pVTZ: Comparison of the
      harmonic frequencies from the TL PhysNet model with the
      reference MP2/aug-cc-pVTZ values. Here, $\Delta\omega =
      \omega_{\rm PhysNet} - \omega_{\rm MP2}$. The average difference
      is 15.5 cm$^{-1}$. This is a comparable performance as for the
      base and refined models at the MP2/6-31G(d,p) level of theory.}
    \label{fig:tl_omega}
\end{figure}

\begin{figure}
    \centering
    \includegraphics[width=0.8\textwidth]{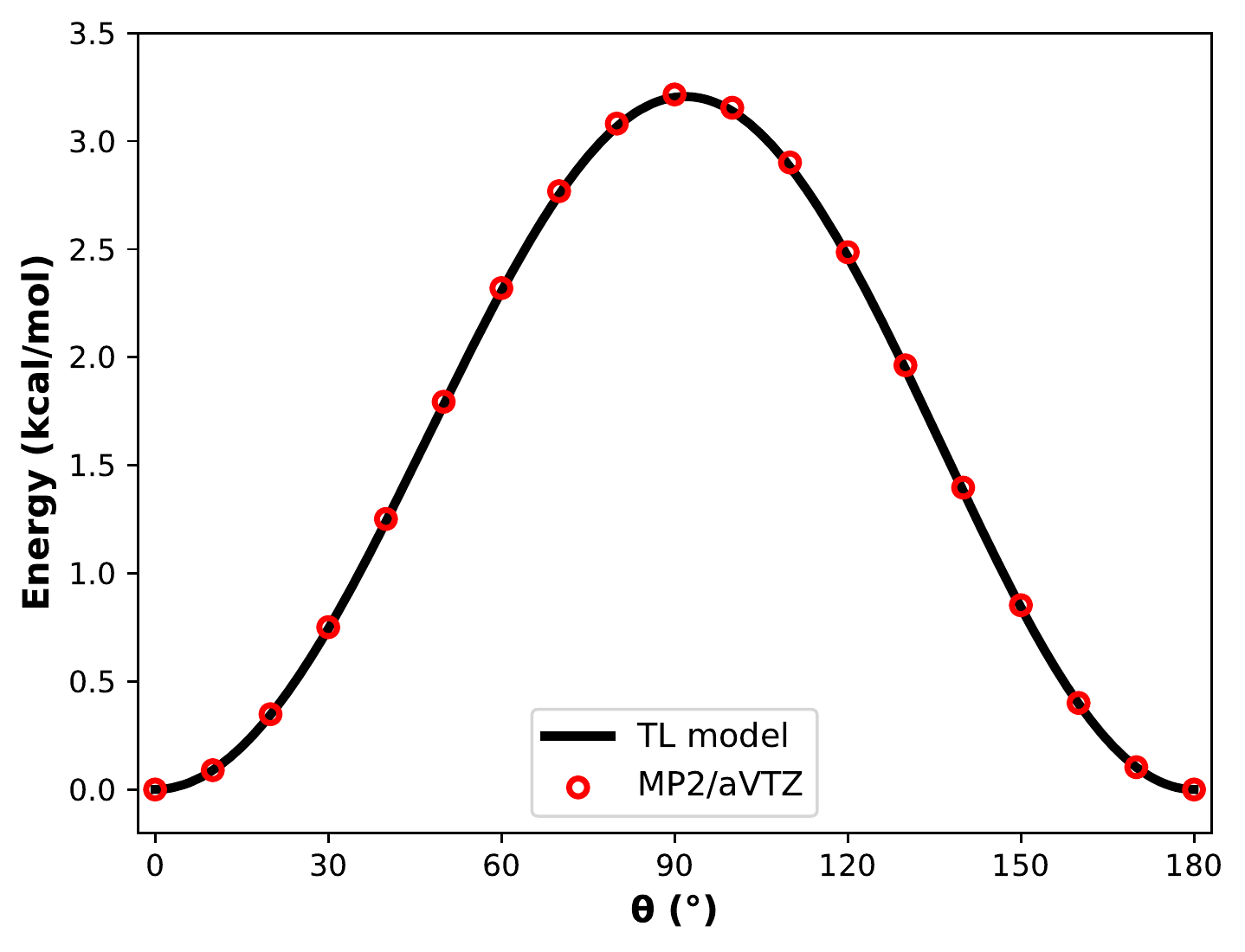}
    \caption{The energy profile of the OH torsion of the
      para-Cl-PhOH. Here, the black solid line represents the energies
      from the TL PhysNet model. The red open circles correspond to the
      reference energies at MP2/aug-cc-pVTZ level of theory.}
    \label{fig:tf_scan}
\end{figure}

\noindent
For the OH-torsional motion the performance of the transfer learned model is
excellent, see Figure \ref{fig:tf_scan}. This training also benefits
from information along the torsional MEP which was included in the
data set explicitly.\\

\section{ML/MM-MD Simulations for Para-Chloro-Phenol}
Next, MD simulations for one para-Cl-PhOH molecule in a box of 881
water molecules were performed using the ML/MM approach. The
intramolecular energy for para-Cl-PhOH atoms was computed by one of
the trained (base, refined) PhysNet models and the TIP3P
model\cite{TIP3P-Jorgensen-1983} was used for water. The internal
structure of each TIP3P water molecule was constrained by the SHAKE
algorithm\cite{shake77}. The electrostatic interactions between
para-Cl-PhOH and water use the fluctuating, geometry-dependent point
charges for the solute together with fixed charges for the
solvent. For the van-der-Waals interactions the Lorentz-Berthelot
combination rules are employed together with Lennard-Jones parameters
$\epsilon$ and $R_\mathrm{min}$ from CGenFF for the
solute.\cite{cgenff2012}\\

\noindent
Initially, the structure of the solvated para-Cl-PhOH molecule in the
periodic water box was relaxed by the steepest descent algorithm for
100 optimization steps. Note that initial atomic structures may be far
away from the equilibrium structure and, thus, large forces act on the
atoms. In such cases, the structures may get distorted to a degree
which is not covered by the reference data set and reactive potentials
such as PhysNet yield wrong energies and forces that even lead to
chemically meaningless conformations. This affects and possibly
invalidates subsequent simulations. One way to avoid this is to use
empirical force fields (with fixed bond declaration) for an initial
structure optimization and it is important to visually check the
optimization results before launching finite-temperature MD
simulations.\\

\noindent
The partially relaxed system was heated to the target temperature of
300\,K in a {\it NVT} simulation for a total of 50\,ps with $\Delta t
= 0.5$\,fs. This was followed by an equilibration simulation of a {\it
  NPT} ensemble for 50\,ps at 300\,K and constant normal pressure with
$\Delta t = 0.2$\,fs. Production simulations were run in the {\it NPT}
ensemble for $1.0$\,ns with $\Delta t = 0.2$\,fs. To validate the
correct implementation of the potential model additional constant
energy simulations $(NVE)$ were performed for 10\,ps and $\Delta t =
0.2$\,fs with initial condition from the final frame of the heating
simulation. The standard deviation of the total energy in the $NVE$
simulations was $< 0.007$\,kcal/mol computed from 500 energy
evaluations every 20\,fs which established conservation of total
energy.\\

\noindent
Infrared (IR) spectra from the simulation data were computed from the
molecular dipole moment time series $\vec{\mu}(t)$ of
para-Cl-PhOH. The molecular dipole moment $\vec{\mu}(t)$ is defined by
the sum of products between each atom position $\vec{x}_i$ and
its respective point charge $q_i$.
\begin{equation}
    \vec{\mu}(t) = \sum_{i}^{N_\mathrm{atoms}} \vec{x}_i \cdot q_i
\end{equation}
The line shape of the IR spectra $I(\nu)$ as a function of the
frequency $\nu$ was obtained by Fourier-transforming the dipole-dipole
autocorrelation function $\left \langle \vec{\mu}(t) \cdot
\vec{\mu}(0) \right \rangle$ and scaled by a quantum correction
factor,\cite{marx:2004} $Q(\nu) = \tanh{(\beta \hbar \nu / 2)}$.
\begin{equation}
  I(\nu) n(\nu) \propto Q(\nu) \cdot \mathrm{Im}\int_0^\infty
  dt\, e^{i\nu t} 
  \sum_{i=x,y,z} \left \langle {\mu}_{i}(t)
  \cdot {{\mu}_{i}}(0) \right \rangle
\label{eq:IR}
\end{equation}
This procedure yields correct lineshapes but not absolute intensities.
For direct comparison, individual spectra were thus multiplied with a
suitable scaling factor $n(\nu)$ to bring intensities of all
spectra to comparable scales.\\

\begin{figure}
    \centering
    \includegraphics[width=0.75\textwidth]{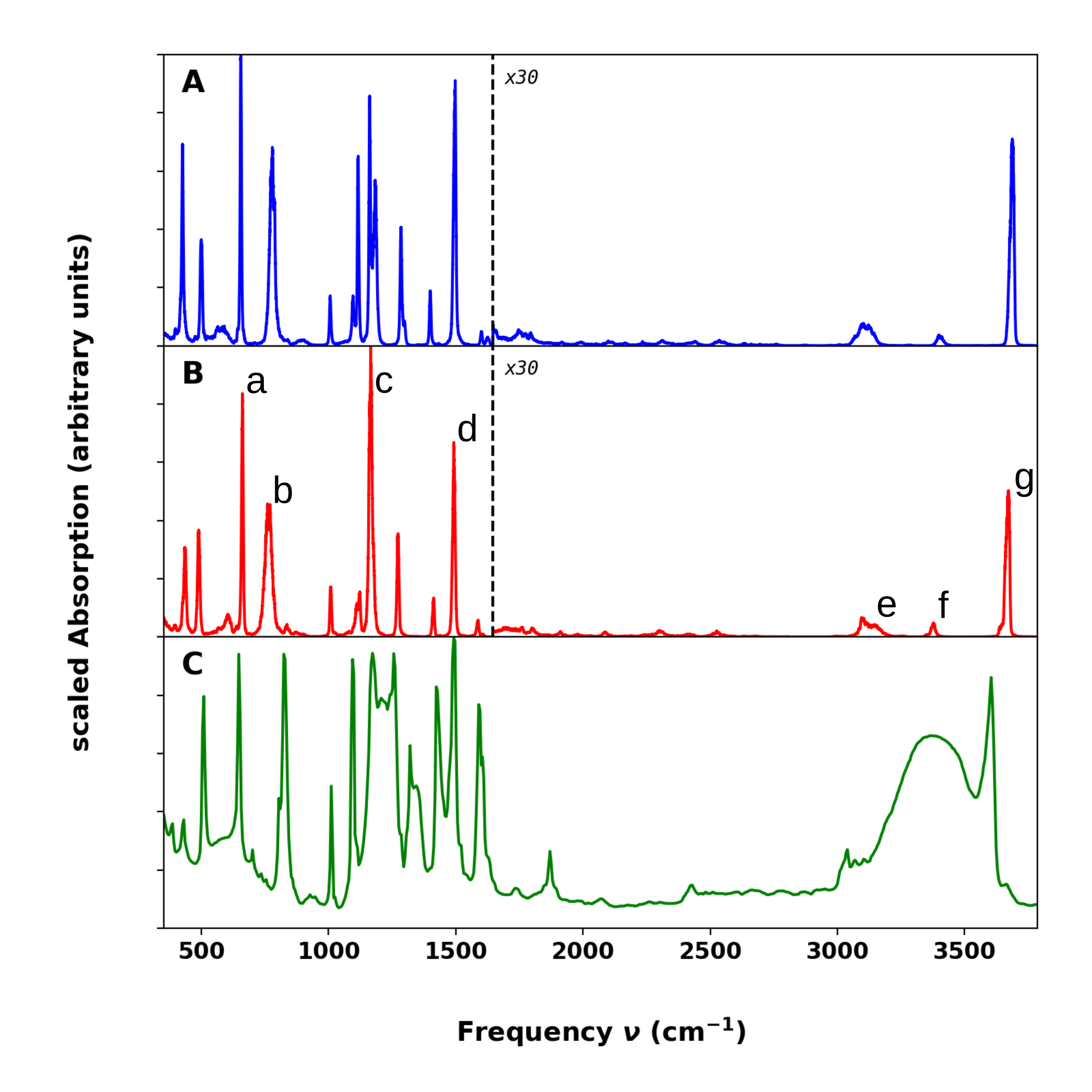}
    \caption{Simulated IR spectra from MD simulation of para-Cl-PhOH
      in water using the (A) PhysNet base model and (B) refined model
      from adaptive training. Absorption intensities for frequencies
      larger than 1720 cm$^{-1}$ (marked by the vertical dashed line)
      are magnified by a factor of 30. Small letters a-g in Panel B
      are labels of selected vibrational modes. Panel C:
      Experimentally measured IR absorption spectra of para-Cl-PhOH
      (10\,wt\%) dissolved in CCl$_4$ (with contaminating water, see
      broad absorption below 3500 cm$^{-1}$).\cite{nistIR}}
    \label{fig:irspec}
\end{figure}

\noindent
Figure \ref{fig:irspec} shows the computed IR spectra from MD
simulation of para-Cl-PhOH in water (panel A and B) and an
experimentally measured IR spectra of a 10wt\% solution of
para-Cl-PhOH in CCl$_4$. Panels A and B show the IR spectra obtained
from running MD simulations using the base and refined model,
respectively. In both cases, the absorption intensities of the IR
spectra for frequencies above $1720$\,cm$^{-1}$ are scaled by a factor
of 30 to improve the visibility of high-frequency signals in the
simulated spectra.\\

\noindent
The positions of the most intense peaks in the computed IR spectra
with the PhysNet base and refined models in Figure \ref{fig:irspec}A
and B remain essentially unchanged. One noticeable exception is signal
{\bf c} that shows the merging of the two peaks in Panel A to a single
peak in Panel B. According to the normal modes of para-Cl-PhOH signal
{\bf c} originates from the C--O stretch and C--OH bend mode. The next
most intense signals {\bf a} and {\bf d} in the IR spectra belong to
the C--Cl stretch and {\bf a} phenol carbon lattice mode,
respectively. Signal {\bf b} can be assigned to a collective
out-of-plane C--H bending mode and has no comparable, experimentally
observed IR signal in Panel C (experiment in CCl$_4$) within its
signal width. Literature reports signal {\bf b} at a higher frequency
above 800\,cm$^{-1}$ for which a corresponding IR signal in Panel C is
observed.\cite{zeegers:2000} The collective four C--H and the single
O--H stretch vibration correspond to signals {\bf e} and {\bf g} in
Figure \ref{fig:irspec}A and B, respectively, are blue-shifted
compared with the experimental IR spectra in panel C. Due to
anharmonicity the frequencies of signals e and g are shifted to the
red compared with the harmonic frequencies of gas phase para-Cl-PhOH
in Figure \ref{fig:freq}. However, the blue shift of computed line
positions even from finite-temperature MD simulations compared with
experiments for high-frequency (X--H) modes is a known deficiency of
using classical MD simulations.\cite{yu:2019} Even with ring polymer
MD simulations the positions of these bands are shifted to higher
frequencies.\cite{yu:2019,MM.oxa:2017} Signal {\bf f} in the computed
IR spectra in panel A and B overlaps in frequency with the broad
signal in panel C around 3400\,cm$^{-1}$, which corresponds to the
stretch vibrations of water contaminating the sample. However, because
water is rigid in the present simulations, the origin of feature {\bf
  f} in the computed spectra remains uncertain. One possibility,
supported by VPT2 calculations not discussed in detail, is the
assignment of peak {\bf f} to a combination band involving a
low-frequency deformation mode of the molecule and the high frequency
C-H stretching modes.\\

\noindent
Contrary to the line positions, more significant differences are found
for the relative intensities of the lines from simulations using the
base and refined PESs. This difference arises because of the
fluctuating charges for the two models differ. While the ratio between
the intensities of the lower frequency bands are qualitatively close
to the experimentally observed ones, the IR signals of the
high-frequency C--H and O--H vibrations (see Figure \ref{fig:irspec}
{\bf e} and {\bf g}, respectively) show much reduced intensities than
in the experiments (note the scaling factor for the simulated IR
spectra). This deficiency very likely originates from the classical
dynamics approach and in particular the zero-point energy
leakage.\cite{sewell:1996} In classical trajectories, it is observed
that energy from vibrational modes of higher frequencies can flow to
lower-frequency modes leading to smaller amplitudes of high-frequency
vibrational modes and reduced intensities in the computed IR spectra
line shape.\\

\begin{figure}
    \centering
    \includegraphics[width=0.75\textwidth]{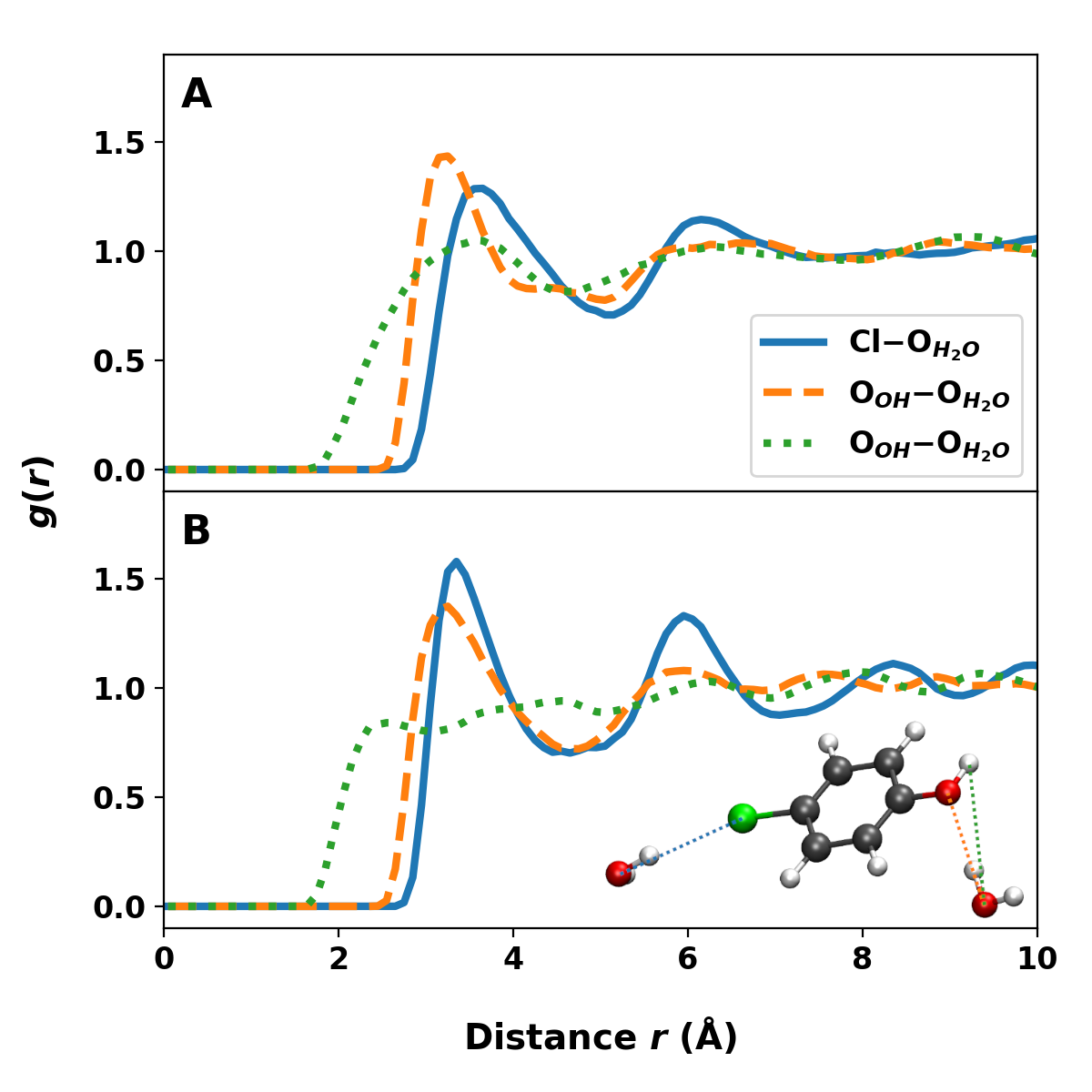}
    \caption{Radial distribution function $g(r)$ between selected
      atoms of para-Cl-PhOH and the oxygen atom of water solvent
      molecules obtained by MD simulation of para-Cl-PhOH in water
      using the (A) PhysNet base model and (B) refined model from
      adaptive sampling. The sketch in panel B visualizes the
      respective atom pairs by the dotted lines.}
    \label{fig:gr}
\end{figure}

\noindent
Figures \ref{fig:gr}A and B show the radial distribution function
$g(r)$ between the Cl, the hydroxyl oxygen and the hydrogen atoms of
para-Cl-PhOH with the oxygen atoms of water from MD simulation using
the PhysNet (A) base and (B) refined model. The amplitudes of the
equivalent $g(r)$ differ between the two PESs as their prediction of
atomic charges differs and, as a consequence, the electrostatic
interactions between para-Cl-PhOH and water changes. Such differences
in the atomic charges between the base and refined PESs also cause the
intensities between the computed IR spectra in Figures
\ref{fig:irspec}A and B to vary.\\

\begin{figure}
    \centering
    \includegraphics[width=0.75\textwidth]{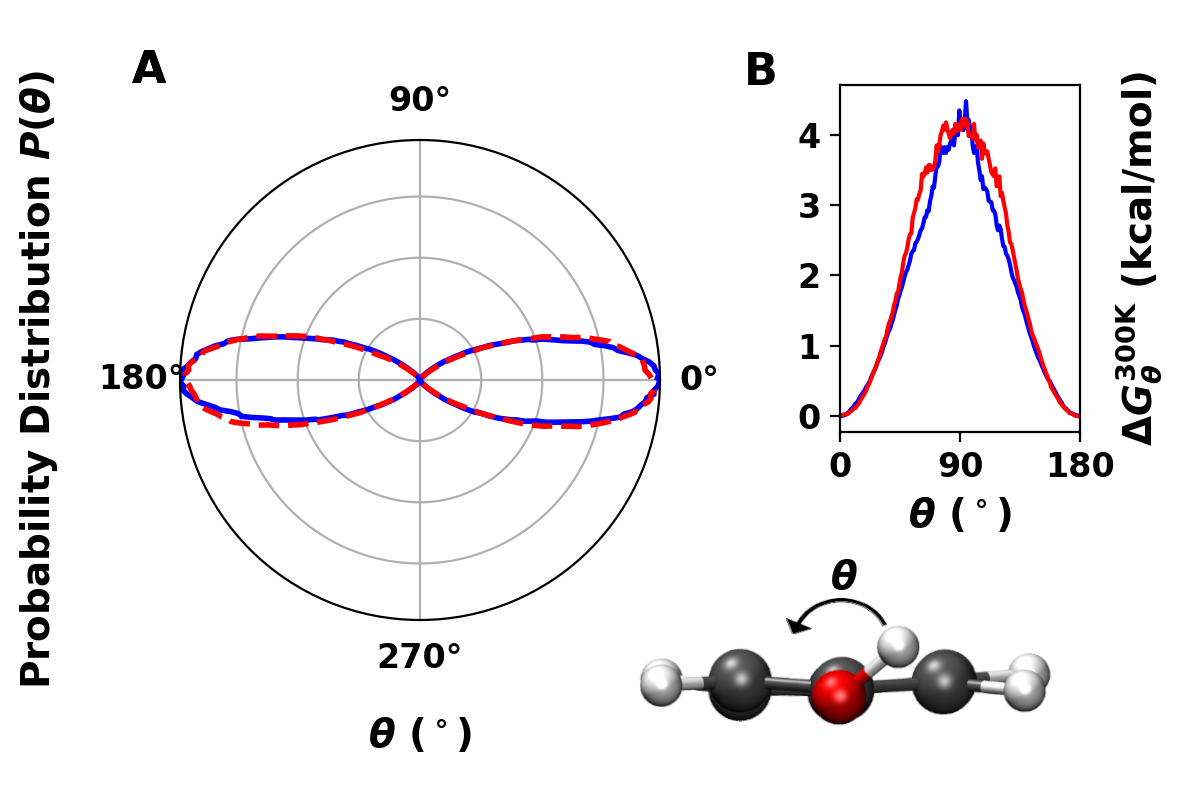}
    \caption{Polar plots of the OH torsion angle probability
      distribution $P(\theta)$ of para-Cl-PhOH in MD simulation using
      the PhysNet base model (blue) and refined model (red) from
      adaptive sampling. Panel B shows the free Energy profile $\Delta
      G(\theta)$ for rotation around the OH torsion angle derived from
      $P(\theta)$ according to $\Delta G(\theta) = -k_B
      T\ln{P(\theta)}$. The barrier height in water is $\sim 4.1$
      kcal/mol compared with $\sim 3.5$ kcal/mol in the gas phase
      which amounts to an increase of $\sim 20$ \% due to the presence
      of solvent.}
    \label{fig:dihe-md}
\end{figure}

\noindent
The probability distribution $P(\theta)$ for the OH torsion angle for
solvated para-Cl-PhOH is shown in Figure \ref{fig:dihe-md}A as a polar
plot with maxima at $0^\circ$ and $180^\circ$ corresponding to planar
conformations using the base (blue) and the refined PES (red
dashed). The free energy profile in Figure \ref{fig:dihe-md}B was
determined from $P(\theta)$ according to $\Delta G(\theta) = -k_B
T\ln{P(\theta)}$ for $T = 300$\,K. Compared with the energy profile of
the OH torsion for para-Cl-PhOH in gas phase (Figure
\ref{fig:scan-rot}), the barrier height for a $180^\circ$ rotation of
the OH group is $\sim 4.1$\,kcal/mol which is $\sim 0.6$\,kcal/mol
higher than for the OH torsional barrier in the gas phase. A higher
free energy barrier in solution originates from the formation of
hydrogen bonds between the -OH group of para-Cl-PhOH and water
molecules as the hydrogen bonds are required to break during the rotation along the OH
torsion angle.\\

\noindent
Comparing the computed IR spectra with experiments or radial
distribution function from MD simulation with results derived from
scattering experiments are possibilities to evaluate the accuracy of
the PES and the simulation setup.\cite{mark:1998,smith:2018} Of
particular interest is the evaluation of the intermolecular
interaction potential between the ML atoms of para-Cl-PhOH and the MM
atoms of water. The electrostatic contributions to the interaction
potential is defined by the atomic charge prediction of the PhysNet
model and the atomic charges defined by solvent model potential,
respectively. The van-der-Waals potential contribution can be adapted
by the modification of the Lennard-Jones parameters assigned to the ML
atoms to match properties such as system density, heat capacity or
solvation free energy. Although the charges on the solute respond to
intramolecular geometry changes, the electrostatic potential for
para-Cl-PhOH can still be further improved by using multipolar models
for the
electrostatics.\cite{MM.mtp:2013,Devereux2014,MM.dcm:2017,MM.dcm:2020,MM.dcm:2022}
This has, e.g., been recently done for fluoro-PhOH which only exhibits
a weak sigma-hole along the C-F
bond.\cite{Clark2007sigmahole,MM.fphoh:2022} For Cl-PhOH such effects
will be stronger due to the pronounced sigma-hole on the chloride
atom.\cite{Clark2007sigmahole,MM.chroma:2018}\\

\section{Summary and Outlook}
In summary, the present contribution describes constructing,
validating and using PhysNet PESs for spectroscopic applications. The
individual steps, including reference data generation, base model
fitting, refinement and TL are described, and the performance of the
NN model is illustrated at every step. The procedure described here is
general and can be adapted to other applications and chemical systems.
It is important to re-iterate that these are only general
guidelines. For specific applications, such as photodissociation
reactions\cite{MM.atmos:2020,MM.criegee:2021,westermayr2022deep},
suitable adjustments to the steps outlined in the present work must be
made. These can, for example, include the adjustment of the data
generation procedure (i.e. for reactive PESs the inclusion of
structures for all transition states and
dissociation/isomerization/etc.  pathways is required) and the
simulation protocol. Irrespective of the application, PESs based on
statistical methods (NNs or kernels) need to be thoroughly tested at
every stage of their construction, and they need to be validated for
both, generalization within the range of structures covered
(interpolation) and outside that range (extrapolation).\\

\noindent
The domain of constructing potential energy surfaces using machine
learning is rapidly expanding and encompasses a wide range of data
generation methods and machine learning techniques, each with its own
strengths and limitations. Therefore, only a subset of these approaches
have been discussed in detail in this work and broader discussions can
be found in recent
reviews\cite{behler2015constructing,bowman:2018,unke:2021,MM.rev:2023}. However,
from a practical and software perspective it is important to stress
that integration of ML-based energy functions into general molecular
simulation software, as shown here for combining PhysNet with CHARMM,
provides enormous "added value" to the simulation community.  Although
pyCHARMM has been previously used in combination with ML
potentials\cite{pycharmm:2023}, the generalization introduced here
will allow the routine ML/MM simulations in CHARMM. Other benefits are
that PESs for small to medium-sized molecules for spectroscopy and
thermodynamic applications can include anharmonicities for all
chemical bonds and the couplings between internal degrees of
freedom. This has, e.g., been demonstrated previously for simulations
of double proton transfer in formic acid dimer.\cite{MM.fad.sol:2022}
"Encoding" changing bond strengths and equilibrium bond lengths
depending on changes in the chemical environment within empirical
energy functions is possible, but very cumbersome.\cite{MM.oxa:2017}
Hence, ML-based PESs make the full complexity of chemistry at a
molecular level visible and available in such a ML/MM//MD
simulation.\\

\noindent
For chemical reactions, which was long a domain of {\it ab initio} MD
simulations, ML-based energy functions now provide the means to run
statistically significant numbers of trajectories which was not
possible before.\cite{MM.criegee:2021,gerber:2018} More recent
applications in this area include malonaldehyde in the gas
phase\cite{schutt:2019,kaser2022transfer}, double proton transfer in
hydrated formic acid dimer\cite{MM.fad:2022}, for atmospherically
relevant reactions using permutationally invariant polynomials and
NN-based energy
functions,\cite{lester:2016,MM.atmos:2020,MM.criegee:2021} or
photochemical reactions.\cite{westermayr2022deep} It will be
interesting to see whether and how these examples, which primarily
concern reactions in the gas phase, can be translated to even more
complex situations such as covalent protein-ligand binding reactions
for which so far primarily studies on diatomic molecules based on
reproducing kernel Hilbert spaces
exist.\cite{MM.mbno:2015,MM.mbno:2016,MM.trhbn:2018}\\

\noindent
Further technical developments are still possible to allow the
integration of other ML potentials as ANI\cite{ani:2017} or
SchNet\cite{schnet:2018} following a similar strategy as the one
described here. In the near future, it is desirable that the
integration of these ML potentials is done directly in the main CHARMM
code to improve the performance and reliability of the simulations. We
plan to make further developments to the PhysNet code to allow the
user to perform the steps described in an automated fashion,
simplifying the setup and simulation of more complex and diverse
systems.\\

\section*{Data Availability Statement}
All scripts used for data generation, training, evaluation and use of
PhysNet will be available at
\url{https://github.com/MMunibas/physnet-pycharmm}.

\bibliography{refs.clean}

\end{document}